\documentclass[12pt,aps,prx,pacs,preprent]{revtex4}
\usepackage{amsmath}
\usepackage{amssymb}
\usepackage{graphicx}
\usepackage{amsfonts}
\usepackage{bm}
\usepackage[latin9]{inputenc}
\usepackage{upgreek}
\usepackage{mathrsfs}
\usepackage{esint}
\usepackage{subfigure}
\usepackage{lineno}
\usepackage{color}
\usepackage{times}

\makeatletter

\def\bra#1{\langle#1|}
\def\ket#1{|#1\rangle}

\makeatother

\begin{document}


\title{Compass-free migratory navigation}

\author{Xin Zhao$^1$, Hong-Bo Chen$^2$, Li-Hua Lu$^{1 \ast}$ and You-Quan Li$^{1,3 \ast}$}

\affiliation{
$^{1}$Zhejiang Province Key Laboratory of Quantum Technology $\&$ Device and\\
Department of Physics, Zhejiang University, Zheda Road 38, Hangzhou 310027, P. R. China\\
$^{2}$School of Information Science and Engineering, NingboTech University, Ningbo 315100, P. R. China\\
$^{3}$Collaborative Innovation Center of Advanced Microstructure,
Nanjing University, Nanjing 210008, P. R. China
   \\
$\ast$To whom correspondence should be addressed; E-mail:  yqli@zju.edu.cn; lhlu@zju.edu.cn
}

\date{\today}


\baselineskip24pt

\begin{abstract}
\begin{center}{\large Abstract}\end{center}
\baselineskip20pt

How migratory birds can find the right way in navigating over thousand miles is an intriguing question,
which much interested researchers in both fields of biology and physics for centuries.
There several putative proposals that sound intuitively plausible all remain contested so far because those hypothesis-models of magnetoreceptor to sense geomagnetic field need either extremely high sensitivity or humankind-like intelligence to guide.
Here we explore theoretically that the birds can navigate to their destination through an entirely new scenario to sense the geomagnetic field.
Our proposal is based on separate peaks of the resonance-fluorescence spectrum of a four-level system derived from the ferric sulfide cluster which exists in a protein complex (Drosophila CG8198) of migratory birds.
As the separation of spectral peaks contains information about geomagnetic field at both current location and birthland, the change of such separation cues the bird to choose a right direction to move and double-resonance emerges once arrived the destination.
Our theoretical mechanism can explain previous experiments on the disorientation of migratory birds caused by oscillating magnetic field naturally and more precisely.
This work provides insight to explain migratory navigation and motivates possible manmade practical devices.
 
\end{abstract}

\bigskip



\maketitle

\newpage


\section{Introduction}

Since the first scientific experiment performed by Thienemann~\cite{Thienemann} in 1927,
people have long been intrigued by how migratory birds can sense the direction
and navigate towards their flight destination.
Kramer considered~\cite{Kramer1953} that the birds should first determine
their relative location to the destination,
and then keep the right direction during the flight,
which means a map and a compass are needed~\cite{LohmannKJ2007}.
The compass relying on the earth's magnetic field (geomagnetic field)
has been a common notion
since it was put forward by Wiltschko~\cite{Wiltschko1972}.
There have been various studies~\cite{Freake2006,Johnsen2008,Kishkinev2015} on
searching for the so-called magnetoreceptor as a biocompass, however,
the physics nature underlying the migratory birds' navigation
is yet to be revealed.

Currently, there are several proposals to identify the magnetoreceptor (MagR).
Given that the spin singlet-to-triplet conversion in certain chemical reaction is affected by magnetic field~\cite{Schulten1982},
a radical pair mechanism~\cite{Ritz2000} was presumed to sense the geomagnetic field.
This was widely discussed in a large number of studies~\cite{Ritz2000,Maeda2008,Rogers2009,Cai2012,Lohmann2016,Hochstoeger,Xiao2020},
but is still under debate
because the influence of geomagnetic field on such a conversion is very weak and
spins need to be relatively isolated from thermal effects~\cite{Johnsen2008}.
Imitating to that the human's compass relies on the magnetic properties of iron minerals,
the birds' navigation was thought to be guided by a magnetite
($\mathrm{Fe_3 O_4}$) in birds' organs acting as a MagR~\cite{Flei2003,Falken2010,Treiber2012,Eder2012,Mouritsen2018}.
Few years ago, Xie's group~\cite{Xie2016} discovered
in certain migratory birds
a rod-like protein complex (\textit{Drosophila} CG8198) which contains Fe-S cluster, and speculated that the complex may play a role of biocompass.
The estimated contribution of iron atoms~\cite{Meister}
in such a putative biocompass is too small to balance the thermal fluctuations of biological systems.
Actually such controversy arises from a subjective picture assuming the MagR as a compass needle.
Thus it is obligated to arouse for an entirely new sight to explain the navigation ability of migratory birds.
The new mechanistic strategy must overcome the disadvantages in current references:
either the influence of geomagnetic field on the chemical reaction is very weak or the orientation of `compass needle' is overwhelmed by thermal fluctuations.

Here we propose a compass-free mechanism for migratory birds' navigation
that can be guided by sensing the spectrum of resonance fluorescence~\cite{Mollow1969}
rather than looking at a magnetite needle.
The information about the destination and the cue for flight direction are naturally included in the separate spectral peaks that depend on the geomagnetic field.
We will start with an investigation on the role of ferric sulfide in the protein complex~\cite{Xie2016} of migratory bird.
Using microscopic formulation~\cite{Katsura} by considering the electronic structure of a ferric sulfide cluster, we evaluate the electric polarization.
Based on the obtained formula of spin-driven electric polarization of the cluster,
we derive a four-level system characterized by two parameters
that depend on both applied magnetic field and electric field.
After solving and analysing the resonance-fluorescence spectrum of the system,
we perceive an entirely new mechanism for migratory birds' navigation,
a much more objective rather than subjective picture.
Our mechanism can overcome the disadvantages of the mechanisms in current references.
We also investigate the spectrum by numerical simulation via Landau-Lifshitz-Gilbert (LLG) equation,
which  enables us to explain the experimentally observed disorientation effects caused by oscillating magnetic field~\cite{Ritz2009} naturally and more precisely.

\parskip=2pt

\section{The role of ferric sulfide cluster}

\begin{figure*}[t]
\centering
\includegraphics[width=0.78\textwidth=1.88]{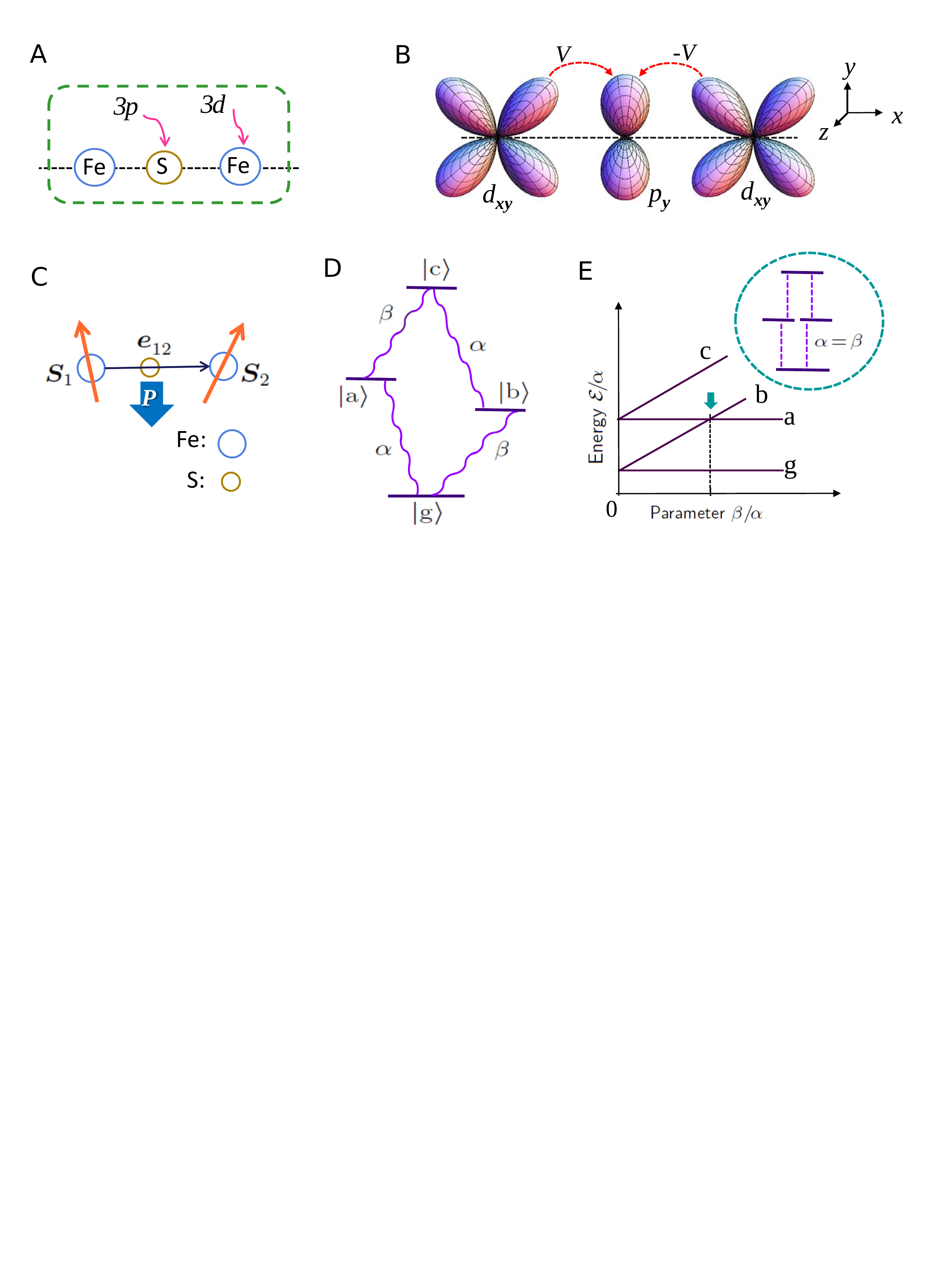}
\caption{\label{fig:model}{\small
\textbf{Illustration for modelling the Fe-S-Fe cluster}.
(\textbf{A}) Schematic of the cluster where ferric ions are bridged by a sulfide ion in between them.
The entanglement between the orbital and spin degree of freedom of the ferric ion is affected by the local magnetic moment arising from its inner-shell $d$-electrons.
(\textbf{B}) Illustration of the Slater-Koster hopping parameters between the active $3d$-orbital and $3p$-orbital of the system.
Its second-order perturbation
hybridizes the $d$- and $p$-orbitals so that an electric polarization is produced.
(\textbf{C}) The spin-driven electric polarization formula in which the induced polarization and spin orders obey certain relation,
$\boldsymbol{P}\propto\boldsymbol{e}^{}_{12}\times(\boldsymbol{S}_1\times\boldsymbol{S}_2)$,
which plays an important role in deriving the four-level system.
(\textbf{D}) Diagram of the derived four-level system that is characterized by two parameters $\alpha$ and $\beta$, i.e.,
$\mathcal{E}_\mathrm{g}=-(\alpha+\beta)/2$,
$\mathcal{E}_\mathrm{a}=(\alpha-\beta)/2$,
$\mathcal{E}_\mathrm{b}=(\beta-\alpha)/2$,
and
$\mathcal{E}_\mathrm{c}=(\alpha+\beta)/2$.
(\textbf{E}) The energy level versus the parameter ratio $\beta/\alpha$,
the four-level model reduces to a three-level one when $\alpha=\beta$.}}
\end{figure*}

Using the method ever applied by Katsura et al. \cite{Katsura} to manganese oxide system,
we consider a three-site cluster (Fig.~\ref{fig:model}A)
where the ferric ion is supposed in an octahedral ligand field so that
the five-fold $d$-orbitals will split into $\mathrm{e}_{\mathrm{g}}$
and $\mathrm{t}_{\mathrm{2g}}$ orbitals.
The $\mathrm{t}_{\mathrm{2g}}$ orbitals, namely $d_{xy}$, $d_{yz}$, and $d_{zx}$
possess lower energy than $\mathrm{e}_{\mathrm{g}}$ orbitals do.
If the on-site spin-orbit interaction is taken into account,
the $\mathrm{t}_{\mathrm{2g}}$ orbitals are entangled with each other
and give rise to the two-fold degenerate states,
\begin{align*}
|\mathrm{I}\rangle=(|d_{xy,\uparrow }\rangle
+|d_{yz,\downarrow }\rangle +i|d_{zx,\downarrow }\rangle )/\sqrt{3},
  \\
|\mathrm{I\!I}\rangle =(|d_{xy,\downarrow }\rangle
-|d_{yz,\uparrow }\rangle +i|d_{zx,\uparrow }\rangle )/\sqrt{3},
\end{align*}
the conventionally so-called $\Gamma _{7}$ states.
The effective exchange interaction reads
$H_{U}=-U\sum_{j}\boldsymbol{\sigma}_{j}\cdot \boldsymbol{S}_{j}$
in which $U$ is the energy of Coulomb repulsion,
$\boldsymbol{\sigma}_{j}$ stands for the Pauli matrix of electron spin, $\boldsymbol{S}_{j}=S\boldsymbol{n}_{j}$ with
$
\boldsymbol{n}_{j}=(\cos \phi _{j}\sin \theta _{j},\sin \phi _{j}\sin \theta
_{j},\cos \theta _{j})
$
for the orientation of the local magnetic moment
that arises from the inner-shell $3d$ electrons in ferric ion
at $j$-th site ($j=1, 2$).
This $H_{U}$ spanned by the aforementioned states $|\mathrm{I}\rangle $ and $|\mathrm{I\!I}\rangle $
can be expressed by a two by two matrix,
which gives us two eigenstates,
$
|\mathrm{P}_{j}\rangle =\sin {\frac{\theta _{j}}{2}}|\mathrm{I}\rangle +\mathrm{e}
^{i\phi _{j}}\cos {\frac{\theta _{j}}{2}}|\mathrm{I\!I}\rangle
$
and
$
|\mathrm{AP}_{j}\rangle =\cos {\frac{\theta _{j}}{2}}|\mathrm{I}\rangle -\mathrm{e}
^{i\phi _{j}}\sin {\frac{\theta _{j}}{2}}|\mathrm{I\!I}\rangle
$
with eigenvalues $E_{|\mathrm{P}_{j}\rangle }=-U/3$ and
$E_{|\mathrm{AP}_{j}\rangle }=U/3$, respectively.
Here $|\mathrm{P}_{j}\rangle $
and $|\mathrm{AP}_{j}\rangle $ are
the states with spin being parallel and anti-parallel to the unit vector $\boldsymbol{n}_{j}$.
The $U$ is regarded as much larger than the other relevant energy scales so that the unperturbed energy
of $|\mathrm{P}_{j}\rangle $ is lower than that of $|\mathrm{AP}_{j}\rangle $.
Therefore we consider only the low-energy manifold spanned by the states
$|\mathrm{P}_1\rangle$ and $|\mathrm{P}_2\rangle$.
For convenience, we can further write states
$|\mathrm{P}_{j}\rangle $ in terms of $d$-orbitals explicitly as
$|\mathrm{P}_{j}\rangle =\sum\limits_{\mu, \sigma}A_{(j)}^{\mu ,\sigma }|d_{\mu
,\sigma }^{(j)}\rangle $,
where $\mu =xy, yz, zx, \,\sigma =\uparrow, \downarrow $,
and the coefficients
$A_{(j)}^{\mu, \sigma }$ are given in supplemental material~\cite{SM} (see Eq.~S5).

Next, we need take the contributions of both ferric and sulfide into account
and hence have a Hamiltonian to describe the on-site energy for
$|\mathrm{P}_{j}\rangle $ of ferric ions and the $3p$-orbitals of sulfide ion,
$
H_\mathrm{site}=\sum\limits_{j=1, 2} E_{|\mathrm{P}_{j}\rangle } c^{\dag}_{|\mathrm{P}_{j}\rangle}
      c^{}_{|\mathrm{P}_{j}\rangle }
 +\sum\limits_{\mu, \sigma } E_{\mathrm{p}}p^{\dag }_{\mu, \sigma } p^{}_{\mu, \sigma }\,
$
where $c^{}_{|\mathrm{P}_{j}\rangle }$ annihilates a spin
parallel state $|\mathrm{P}_{j}\rangle $ at site $j$,
and $p^{}_{\mu,\sigma }$ with $\mu=x, y, z$ annihilates the one for the $3p$-orbitals of sulfide ion,
$E_{\mathrm{p}}$ is the energy of $p$-orbitals
and the subscript $\sigma$ denotes the spin label.
The hybridization between the sulfide's  $p$-orbitals and spin parallel states
$|\mathrm{P}_{j}\rangle $ of ferric ion is governed by the hopping between them, namely
$$
H_{V}=V\sum_{\sigma, j}(-1)^{j-1}\bigl( p_{y,\sigma }^{\dagger }d_{xy,\sigma
}^{(j)}+p_{z,\sigma}^{\dagger}d_{zx,\sigma}^{(j)}+\mathrm{h.c.}\bigr),
$$
where $d_{\mu ,\sigma }^{(j)}$ ($\mu =xy, zx$) are annihilation operators for
$3d$ states of ferric ions.
The superscript $j=1, 2$ in $d_{\mu,\sigma }^{(j)}$ denotes the site label.
The $V=t_{pd\pi}$ is the hopping strength between the ferric ion and the sulfide ion,
which is determined by the Slater-Koster's rules~\cite{Slater} (Fig.~\ref{fig:model}B).
The bases of the Hilbert space of our cluster model
consist of the two-fold degenerate states
$|\mathrm{P}_{1}\rangle $, $|\mathrm{P}_{2}\rangle $,
as well as $p_{\mu ,\sigma }$ ($\mu=y, z,$ $\sigma=\uparrow, \downarrow $),
thus a perturbed eigenstate is a linear superposition of these states.
Taking the Hamiltonian $H_{V}$ as a perturbation and
using the second-order perturbation theory~\cite{Landau}
within the two-fold degenerate states,
we obtain eventually two eigenstates,
$|\psi _{-}\rangle $ and
$|\psi _{+}\rangle $.

Then the electric polarization produced by aforementioned cluster is defined by
$\boldsymbol{P}=\left\langle e\boldsymbol{r}%
\right\rangle $ in which
 $\boldsymbol{r}$ is the position operator and $e$ the elementary
electric charge.
Let us consider superexchanges where two holes
are put into the ground state $|\psi _{-}\rangle $ and the next low-lying
state $|\psi _{+}\rangle $,
the electric dipole moment is hence obtained
\begin{equation}
\boldsymbol{P}=
P \,\boldsymbol{e}^{}_{12}\times
  (\boldsymbol{S}_1\times \boldsymbol{S}_2),
\label{eq:polar}
\end{equation}
where $\boldsymbol{e}_{12}^{{}}$ denotes the unit vector parallelling to the
direction of the bond from ferric ion at site-$1$ to that at next site-$2$,
and
$P=-4e\left(V/\Delta\right)^3 I/9$
with $I$ being the overlap integral,
$
I=\int\mathrm{d}^{3}\boldsymbol{r}d_{yz}(\boldsymbol{r}) y\, p^{}_{z}(\boldsymbol{r}+\boldsymbol{r}_0)
$
in which $\boldsymbol{r}_{0}$ is the inter-atomic distance between ferric
and sulfide ions.
We have taken the approximation
$
\left\langle \psi _{\mp }|\psi _{\mp }\right\rangle ^{-1}
 \approx 1- 2(V/\Delta )^{2} (1\pm \left\vert \eta\right\vert)/3$
in deriving to Eq.~(\ref{eq:polar}).
Now we visualize (Fig.~\ref{fig:model}C) the relationship
between the spin order ($\boldsymbol{S}_1$, $\boldsymbol{S}_2$) and the induced electric polarization $\boldsymbol{P}$.
Consequently, the electric polarization can be expressed in terms of two neighbor-site spin operators~\cite{Katsura}.
We employ a typical set of parameters~\cite{Clementi}:
$e=1.6022\times 10^{-19} \mathrm{C}$, $V=0.75\,\mathrm{eV}$, $\Delta =3\,\mathrm{eV}$, $Z_{\mathrm{S}}=5.48$,
$Z_{\mathrm{Fe}}=11.18$,
$r^{}_0=4.2630\,a^{}_0$  for numerics and obtain
$I\approx -0.0133 a^{}_0$ and
$P=7.8312\times 10^{-34}\,\mathrm{Cm}$.

\section{Modelling to a four-level system}

Now we derive the modelling Hamiltonian of the aforementioned Fe-S-Fe cluster
if both a magnetic field $\boldsymbol{B}$ and an electric field $\boldsymbol{E}$ are applied respectively,
\begin{align}
\mathscr{H}=-\boldsymbol{B}\cdot(\boldsymbol{S}_1+\boldsymbol{S}_2)
-\boldsymbol{E}\cdot\bigl(\boldsymbol{e}^{}_{12}\times(\boldsymbol{S}_1 \times\boldsymbol{S}_2)\bigr).
\label{eq:ZemmanStark}
\end{align}
For simplicity, we have set the units of both the magnetic and electric fields by taking account of coefficients from the magnetic moment and that from the electric polarization.
We introduce the total spin operator $\boldsymbol{S}=(\boldsymbol{S}_1+\boldsymbol{S}_2)$,
a new operator $\boldsymbol{Q}= 2 \boldsymbol{S}_1\times\boldsymbol{S}_2$
as well as a vector field
$\boldsymbol{C}=\frac{1}{2}(\boldsymbol{E}\times\boldsymbol{e}^{}_{12})$.
These operators fulfill the following commutation relations
\begin{eqnarray*}
 & &\bigl[ S_i,\,S_j \bigr]=i\epsilon^{}_{ijk}S_k,
   \,\,\,
 \bigl[Q_i,\, S_j\bigr]=i\epsilon^{}_{ijk}Q_k,
   \\
 & &\bigl[Q_i,\, Q_j\bigr]=i\epsilon^{}_{ijk}S_k,
\end{eqnarray*}
where the subscripts denote $x, y, z$.
One can verify that further combinations,
$J_i=(S_i + Q_i)/2$ and
$K_i=(S_i - Q_i)/2$
become two sets of decoupled SU(2) generators.
Thus our modelling Hamiltonian can be expressed as
\begin{equation}
\mathscr{H}=-\boldsymbol{\alpha}\!\cdot\!\boldsymbol{J}-\boldsymbol{\beta}\!\cdot\!\boldsymbol{K}.
\label{eq:4state-model}
\end{equation}
in which two vector fields $\boldsymbol{\alpha}=(\boldsymbol{B}+\boldsymbol{C})$ and
$\boldsymbol{\beta}=(\boldsymbol{B}-\boldsymbol{C})$ parameterize the model.

To solve the eigenenergies and eigenstates of the SU(2)$\times$SU(2) Hamiltonian Eq.~(\ref{eq:4state-model}),
it is more convenient to choose the direction of $\boldsymbol{\alpha}$ as the $z$-axis of our coordinate frame.
Then it can be simplified and solved by four eigenvectors $\ket{\mathrm{g}}$, $\ket{\mathrm{a}}$, $\ket{\mathrm{b}}$ and $\ket{\mathrm{c}}$ ~\cite{SM} (details see Eq.~S19)
with eigenenergies:
$\mathcal{E}_\mathrm{g}=-(\alpha+\beta)/2$,
$\mathcal{E}_\mathrm{a}=(\alpha-\beta)/2$,
$\mathcal{E}_\mathrm{b}=(\beta-\alpha)/2$,
and
$\mathcal{E}_\mathrm{c}=(\alpha+\beta)/2$ correspondingly.
Therefore, we arrive at a four-level system characterized by two parameters.
This model and its energy levels are illustrated in Figs.~\ref{fig:model}D, \ref{fig:model}E,
which reduces to a three-level model (Fig.~\ref{fig:model}E, inset)
at the level-crossing point $\beta/\alpha=1$.
Thus we have a four-level system modeled by following Hamiltonian in Fock space:
\begin{align}
H_\mathrm{FeS}
 = &\mathcal{E}_\mathrm{g} \ket{\mathrm{g}}\bra{\mathrm{g}}
   + \mathcal{E}_\mathrm{a} \ket{\mathrm{a}}\bra{\mathrm{a}}
       \nonumber\\[1mm]
  & + \mathcal{E}_\mathrm{b} \ket{\mathrm{b}}\bra{\mathrm{b}}
 + \mathcal{E}_\mathrm{c} \ket{\mathrm{c}}\bra{\mathrm{c}}.
\label{eq:4level-model}
\end{align}

\section{Optical characteristics}

To investigate optical characteristics of the four-level system derived in previous section,
we start with the total Hamiltonian in Schr\"odinger picture, namely
\begin{equation}
H_\mathrm{tot}=H_\mathrm{FeS}+H_\mathrm{bath}+ H_\mathrm{int}+H_\mathrm{drive},
\label{eq:tot}
\end{equation}
where  $H_\mathrm{FeS}$, $H_\mathrm{bath}$ and $H_\mathrm{int}$
describe the four-level system,
its thermal bath,
and the interaction between them.
Here $H_\mathrm{drive}$ represents the contribution of a classical  driving field applied to the system (Fig.~\ref{fig:spectrum}A).
It will be technically helpful to adopt the rotating frame in accord with the frequency $\nu$ of driving field.
In terms of an appropriate choice of $H_0$~\cite{SM} (see eq.~S27),
we transform the above Hamiltonian Eq.~(\ref{eq:tot}) in interaction picture, consequently,
\begin{eqnarray}
H^\mathrm{I}(t)
 &=& \sum_{a} \Delta_a \ket{a}\bra{a} + \sum_{j,\boldsymbol{k},s}\Bigl[\Omega_j \gamma^{}_j
  \nonumber \\
 & & + g_k^{(j)} a^{\dagger}_{\boldsymbol{k} s} \gamma^{}_j \mathrm{e}^{i(\omega_{\boldsymbol{k} s}-\nu)t}
   + \mathrm{h.c.} \Bigr].
\label{eq:Hamiltonian}
\end{eqnarray}
Here
$\gamma^{}_1=\ket{\mathrm{g}}\bra{\mathrm{a}}+\ket{\mathrm{b}}\bra{\mathrm{c}}$
and
$\gamma^{}_2=\ket{\mathrm{g}}\bra{\mathrm{b}}+\ket{\mathrm{a}}\bra{\mathrm{c}}$
are energy lowering operators,
$\Omega_j$ ($j=1, 2$)
is called Rabi frequency that refers to the strength of driving,
while
$\Delta_{a}$ ($a=\mathrm{a}$, $\mathrm{b}$ and $\mathrm{c}$)
denote the detunings between the level differences and the driving-field frequency, i.e.,
$\Delta_\mathrm{a}=\alpha/\hbar-\nu$,
$\Delta_\mathrm{b}=\beta/\hbar-\nu$,
$\Delta_\mathrm{c}=(\alpha+\beta)/\hbar-2\nu$.
Applying the generalized reservoir theory with Weisskopf-Wigner approximation,
we can derive the equation of motion~\cite{Lindblad1976} for the reduced density matrix remained on the Hilbert space of the four-level system,
\begin{equation}
\frac{\mathrm{d}}{\mathrm{d}t} \rho=\frac{1}{i\hbar}[H,\,\rho] + \mathscr{L}(\rho),
\label{eq:Lindblad}
\end{equation}
where the reduced Hamiltonian reads
$H=\sum_a\Delta_a\ket{a}\bra{a} +\sum_j\Omega_j(\gamma^{}_j+\gamma_j^\dagger)$
and the Lindblad damping part in our model reads
$\mathscr{L}(\rho)=\Gamma_1 L_{11}+\Gamma_2 L_{22} +\Gamma_{12}(L_{12}+L_{21})$
with
$L_{ij}=(2\gamma^{}_i\rho\gamma^\dagger_j-\gamma^\dagger_i\gamma^{}_j\rho-\rho\gamma^\dagger_j\gamma^{}_i)/2$
and
$\Gamma_j$ the natural width of the transitions.

Once explicitly writing out the equation of motion Eq.~(\ref{eq:Lindblad}) in component form,
we get a set of first order differential equations for the reduced density matrix.
Since the trace of the density matrix is always a unit, i.e.,
$\rho_\mathrm{gg}+\rho_\mathrm{aa}+\rho_\mathrm{bb}+\rho_\mathrm{cc}=1 $,
we can substitute $\rho_\mathrm{gg}$ by the other three diagonal elements
and obtain
\begin{equation}
\frac{\mathrm{d}}{\mathrm{dt}}\Psi=M\Psi + \Lambda.
\label{eq:diff-equation}
\end{equation}
Here $M$ is a $15\times 15$ matrix,
$\Psi$ and $\Lambda$ are single column matrices with 15 entries.
The inhomogeneous term $\Lambda$ arises from the aforementioned substitution
and contains 4 non-vanishing entries,
$\Lambda_8=-i\Omega_2$,
$\Lambda_{12}=-i\Omega_1$,
$\Lambda_{14}=i\Omega_2$,
$\Lambda_{15}=i\Omega_1$.
The formal integration of Eq.~(\ref{eq:diff-equation})
can be obtained by making a Laplace transform,
\begin{equation}
\Psi(z)=(z-M)^{-1}\Psi(t_0)+(z-M)^{-1}\frac{\Lambda}{z},
\label{eq:formal}
\end{equation}
where $\Psi(t_0)$ is the integration constant from initial value.

\subsection*{Spectrum of resonance fluorescence}

The emission spectrum is defined as the Fourier transform of the double-time correlation function~\cite{Glauber1963},
$$
\mathcal{S}(\omega)=\pi^{-1} \mathrm{Re}\int^{\infty}_{0}
\left\langle\mathsf{E}^{(-)}(\boldsymbol{r}, t+\tau)
\mathsf{E}^{(+)}(\boldsymbol{r},t)\right\rangle\mathrm{e}^{-i\omega\tau}\mathrm{d}\tau,
$$
of the negative and positive parts of the effective electric field operators
$\mathsf{E}^{(\mp)}$ that can be expressed by effective electric polarization operator
$$
\mathsf{P}^{(-)}(t)=\boldsymbol{d}^{}_1\gamma^\dagger_1(t)+\boldsymbol{d}^{}_2\gamma^\dagger_2(t),
$$
and $\mathsf{P}^{(+)}(t)=\bigl(\mathsf{P}^{(-)}(t)\bigr)^\dagger$.
Here $\boldsymbol{d}_1$ and $\boldsymbol{d}_2$ denote the dipole moments.
The expression Eq.~(\ref{eq:formal}) enables us to evaluate
$\left\langle\mathsf{P}^{(-)}(t_0+\tau)\right\rangle$ and
$\left\langle\mathsf{P}^{(+)}(t_0)\right\rangle$,
furthermore
their double-time correlation function according to the quantum regression theorem~\cite{Lax1963}.
Finally we obtain
\begin{equation}
\mathcal{S}(\omega)=\mathrm{Re}\left\langle\mathsf{P}^{-}(z)\cdot\mathsf{P}^{(+)}(t_0)
\right\rangle_\mathrm{incoh}\Bigr|_{z=i\omega},
\label{eq:spectrum}
\end{equation}
where the details are given in supplemental material (see Eq.~S49).

\begin{figure*}[t]
\centering
\includegraphics[width=0.90\textwidth=1.82]{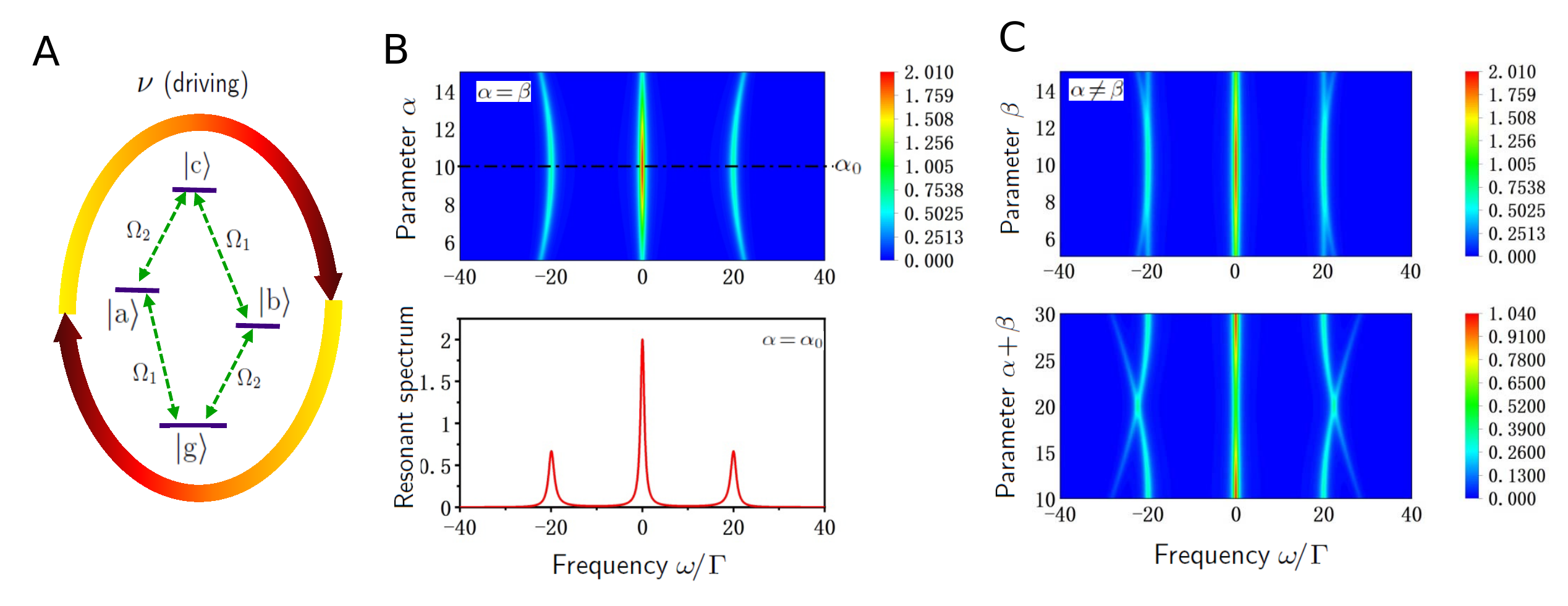}
\caption{\label{fig:spectrum}{\small
\textbf{The feature of resonance-fluorescence spectrum.}
(\textbf{A}) The schematic of the four-level system in the presence of a classical driving field of frequency $\nu$ where the Rabi frequencies $\Omega_1$ and $\Omega_2$ coupling different
states are specified.
The feature of the spectrum depends on two parameters $\alpha$ and $\beta$.
(\textbf{B}) In the regime of $\alpha=\beta$, the spectrum always possesses three peaks.
The bottom panel exhibits the spectral curve at the double-resonance point $\alpha=\beta=\alpha_0$.
(\textbf{C}) When $\alpha\neq\beta$, the spectrum usually contains five peaks:
a main peak at the center and two side peaks symmetrically locating at left- and right-hand sides respectively.
The top panel shows the case that $\beta$ changes from $5$ to $15$ while keeping $\alpha=\alpha_0$ fixed,
which belongs to single-resonance regime.
The five peaks are merged into three peaks
if $\beta$ approaches to the resonance point $\alpha_0$.
The bottom panel manifests the spectra along another line
in which both $\alpha$ and $\beta$ vary while keeping $\beta-\alpha=10$.
In these panels,
the frequency of the main peak is chosen as the origin point of the $x$-axis and
the magnitude in the color-bar measures the height of peaks of the spectrum.
The data for the above figures are  calculated by choosing $\Omega/\Gamma=10$ and $\alpha_0/\Gamma=20$. }}
\end{figure*}

If we let $\alpha=\beta$, those complicated expressions can be simplified to certain extent,
and furthermore under double-resonance conditions
$\Delta_\mathrm{a}=\Delta_\mathrm{b}=0$,
i.e., a particular $\alpha_0=\hbar\nu$,
we can solve an  analytical expression for the spectrum
\begin{equation}
\mathcal{S}(\omega) = R(\omega-\lambda_0) L(\omega-\lambda_{-}) L(\omega-\lambda_{+}),
\label{eq:theSpectrum}
\end{equation}
where
$\lambda_0 =\nu$,
$\lambda_\pm=\lambda_0\pm\sqrt{4\Omega^2-(\Gamma/4)^2}$.
Here the functions are
\begin{align*}
L(x)&=\frac{4\Omega^2}{x^2+(3\Gamma/4)^2},
  \\[1mm]
R(x)&=\frac{\Gamma}{\Gamma^2+8\Omega^2} \frac{x^2+\Gamma^2 + 2\Omega^2}{x^2+(\Gamma/2)^2}.
\end{align*}
In the same regime $\alpha=\beta$ with off double-resonance $\Delta^{}_a\neq 0$, the spectrum still  possesses three peaks (see Fig.~\ref{fig:spectrum}B) although we do not have an analytical expression like Eq.~(\ref{eq:theSpectrum}).

In the other regime $\alpha\neq\beta$, the spectrum usually contains five peaks:
a main peak at the center and two side-peaks symmetrically locating at left- and right-hand side respectively.
When $\beta$ approaches to $\alpha$,
the five peaks merge into a three-peak shape.
The feature of the spectrum relies on parameters $\alpha$ and $\beta$,
of which the rich picture is manifested in Fig.~\ref{fig:spectrum}.
The theory of resonance fluorescence was first put forward by Mollow~\cite{Mollow1969}
for two-level system and the obtained spectrum was then measured in experiment~\cite{Grove1977}.
Later, it was also applied to a three- and a four-level system~\cite{Scully1990,Zhu1996} successively.

\section{Numerical Simulation of the side peaks}\label{sec:LLG}

Let us revisit the Hamiltonian Eq.~(\ref{eq:ZemmanStark}) given previously  for the Fe-S-Fe cluster in the presence of magnetic and electric fields.
We choose the direction of the static magnetic field $\boldsymbol{B}$
as $y$-axis, i.e., $\boldsymbol{B}\parallel\hat{y}$,
and the static electric field $\boldsymbol{E}=E(0,\sin \theta ,\cos \theta )$
within $y$-$z$ plane with an inclination angle $\theta $ to the $z$-axis.
We carry out the numerical simulation of the spin resonance modes by solving the Landau-Lifshitz-Gilbert (LLG) equation in terms of the fourth-order Runge-Kutta method.
The LLG equation is given by
\begin{equation}
\frac{\partial \boldsymbol{S}_i}{\partial t}=-\frac{1}{1+\alpha_{G}^{2}}
 \left[
  \boldsymbol{S}_{i}\!\times\!\boldsymbol{H}_{i}^{\mathrm{eff}}+\frac{\alpha^{}_{G}}{S}%
\boldsymbol{S}_{i}\!\times\!(\boldsymbol{S}_{i}\!\times\!\boldsymbol{H}_{i}^{\mathrm{eff}})%
\right].
\label{LLG}
\end{equation}
where $\alpha^{}_{G}~(=0.01)$ is the Gilbert-damping parameter, and
\[
\boldsymbol{H}_{i}^{\mathrm{eff}}=-\frac{\partial\mathscr{H} }{\partial \boldsymbol{S}_{i}}
\]%
is the effective local field acting on the $i$th spin $\boldsymbol{S}_{i}$.
To study the spin excitation modes according to the approach of Mochizuki~\cite{Mochizuki},
we calculate the dynamical magnetic
susceptibilities defined by
\begin{equation}
\chi _{\mu }^{\mathrm{mm}}(\omega )=\frac{\Delta S_{\mu }(\omega )}{B_{\mu
}(\omega )},
\end{equation}%
where the subscript $\mu $ stands for $x$, $y$ or $z$, and
$B_{\mu }(\omega )$ is the Fourier transform of the time-dependent pulse of magnetic field $B_{\mu }(t)$.
Here, a time-localized, uniform $\delta $-function
pulse: $\,\boldsymbol{B}(t)=\delta (t)\hat{z}$, is applied in the LLG
simulation,
and $\Delta S_{\mu }(\omega )$ is the Fourier transform of the $%
\mu $-component of the net spin: $\Delta \boldsymbol{S}(t)=\boldsymbol{S}(t)-\boldsymbol{S%
}(0)$ with
$\boldsymbol{S}(t)=N^{-1}\sum_{i}\boldsymbol{S}_{i}(t)$, which is the
transient response of the system under the intense pulse of external
magnetic field.

Starting with a random initial spin configuration for  $\boldsymbol{S}
_{1}$ and $\boldsymbol{S}_{2}$, we relax them using LLG-equation simulation by
a sufficient time evolution to obtain the ground state. We then simulate
the spin dynamics by applying an intense pulse of magnetic field to obtain
the spin excitation modes.
We present the calculated time evolution of $z$-component of $\boldsymbol{S}(t)$
(Fig.~\ref{fig:sw}A)
after applying the $\delta $-function magnetic-field pulse along the $z$-axis at $%
t=0$.
We show the calculated imaginary part of
the dynamical magnetic susceptibility Im$\chi _{z}^{\mathrm{mm}}(\omega )$
(Fig.~\ref{fig:sw}B)
via the Fourier transform of $\Delta S_{z}(t)$.
We can see that there are two resonance peaks in the spectra and the peak positions denoting the
spin excitation modes are located at $\omega _{1}=0.93$, and $\omega _{2}=2.124$ (in unit of $\omega^{}_\mathrm{L}$),
respectively.
Considering different geomagnetic field by tuning with appropriate electric field,
we calculate spectra
Im$\chi _{z}^{\mathrm{mm}}(\omega )$ for various $B$ and $E$
by the simulation method (Fig.~\ref{fig:more})
and find the same feature as shown in Fig.~\ref{fig:strategy}C.
The lower-lying resonant modes have
the identical frequency, in consistent with the
spectrum of the resonance fluorescence (Fig.~\ref{fig:strategy}C).

\section{Strategy for navigation}

\begin{figure*}[t]
\centering
\includegraphics[width=0.88\textwidth=1.82]{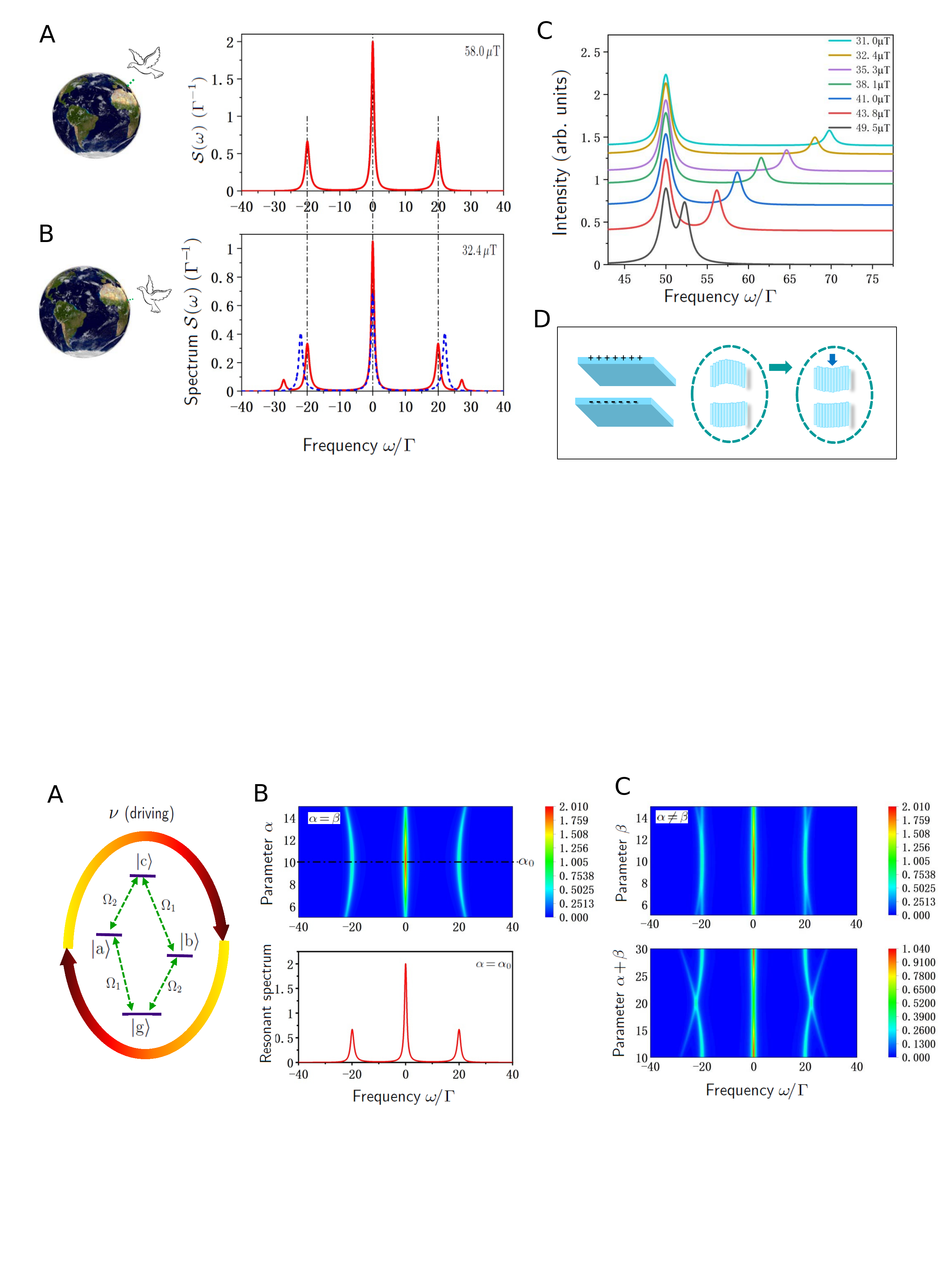}
\caption{\label{fig:strategy}\small
\textbf{Schematic of navigation strategy.}
(\textbf{A}) The spectrum of the resonance fluorescence at the birthland of the migratory bird,
which is under double-resonance condition, $\beta=\alpha=\alpha_0$.
(\textbf{B}) The sensory spectrum of resonance fluorescence (red line) at a place away from the bird's birthland.
The migratory bird is assumed to tune the parameter of the model to match the resonance condition
for $\alpha$ with the help of its internal electric field
although the geomagnetic field at the location $x\neq x_\mathrm{b}$ does not match the resonance condition.
The spectrum without the tuning field processes three peaks merely (blue dashed-line).
The resonance spectrum (red line) tuned up by the internal electric field
contains five peaks, and the splitting side-peak contains information of the location.
The curves are plotted by choosing
$\Omega/\Gamma=10$ and $\alpha_0/\hbar=10.4564\,\mathrm{rad/\upmu s}$
for $B(x_\mathrm{b})=58.0\,\mathrm{\upmu T}$ at $52^\circ$ latitude and $B(x)=32.4 \mathrm{\upmu T}$  near the equator.
(\textbf{C}) Comparison of the side peaks under the single-resonance condition at different places
(parameter choice $\Omega/\Gamma=25$ for a better peak-resolution of more data).
(\textbf{D}) Schematic of how the internal electric field is tuned.
}
\end{figure*}

\begin{figure*}[t]
\centering\includegraphics[width=95mm]{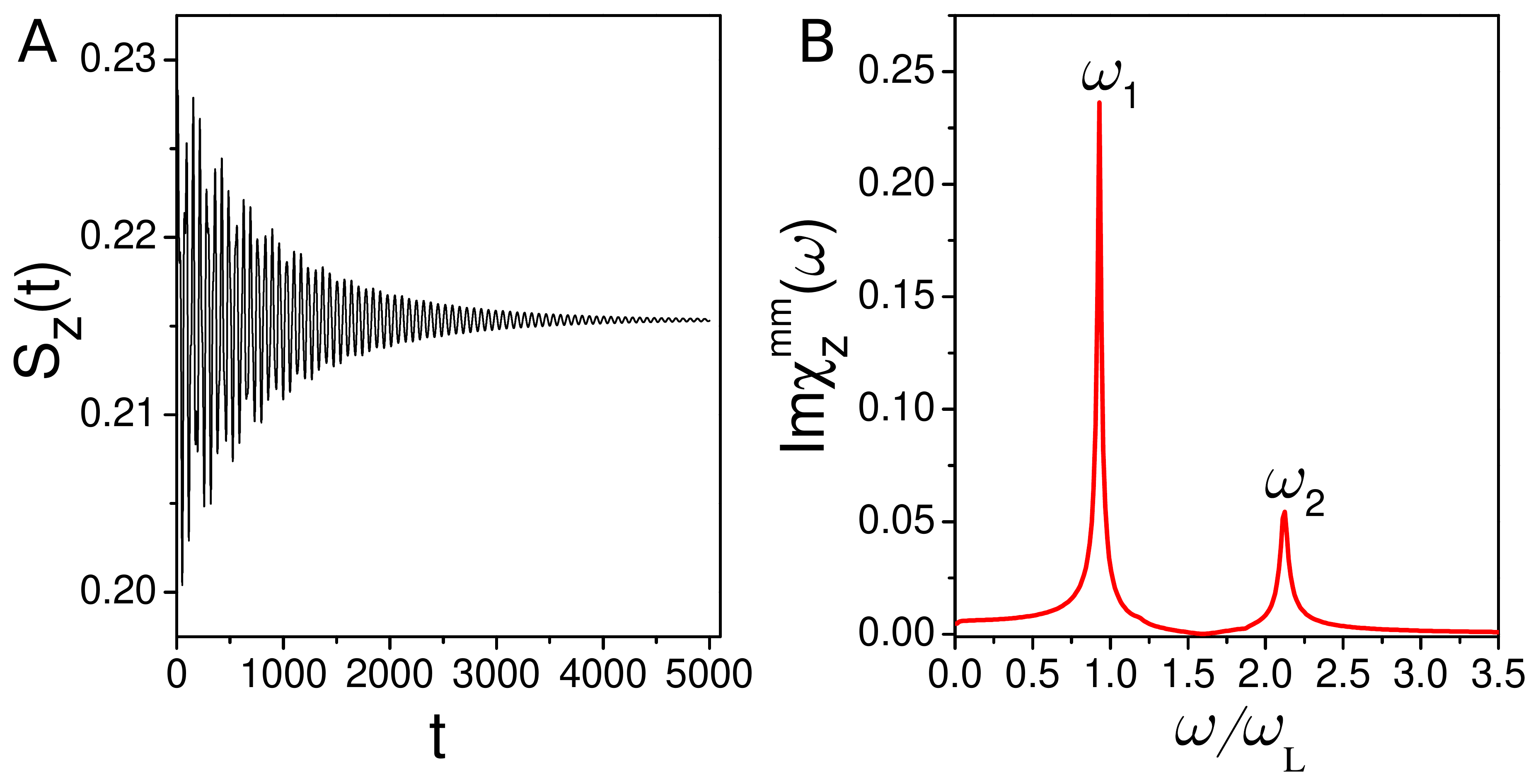}
\caption{\small
\textbf{Simulation of peaks}.
(\textbf{A}) The time evolution of $z$-component of $\boldsymbol{S}(t)$
after applying an intense $\delta$-function magnetic field pulse at
$t=0$.
The parameters are adopted as follows: $B=E=0.1$,
$\protect\theta =\pi /4$, $\alpha^{}_{G}=0.01$.
(\textbf{B}) The imaginary part of the
dynamical magnetic susceptibility $\mathrm{Im}\chi _{z}^{\mathrm{mm}}(\omega )$.
The peaks' positions indicate the frequency of the spin
excitation modes of the Fe-S-Fe cluster. The frequency $\protect\omega $ is measured
in unit of $\protect\omega _\mathrm{L}=\protect g\gamma^{}_{e}B/\hbar$,
with $\protect\gamma_{e}$ the gyromagnetic ratio and $g$ the g-factor of ferric.}
\label{fig:sw}
\vspace{160mm},
\end{figure*}

\begin{figure*}[t]
\centering\includegraphics[width=68mm]{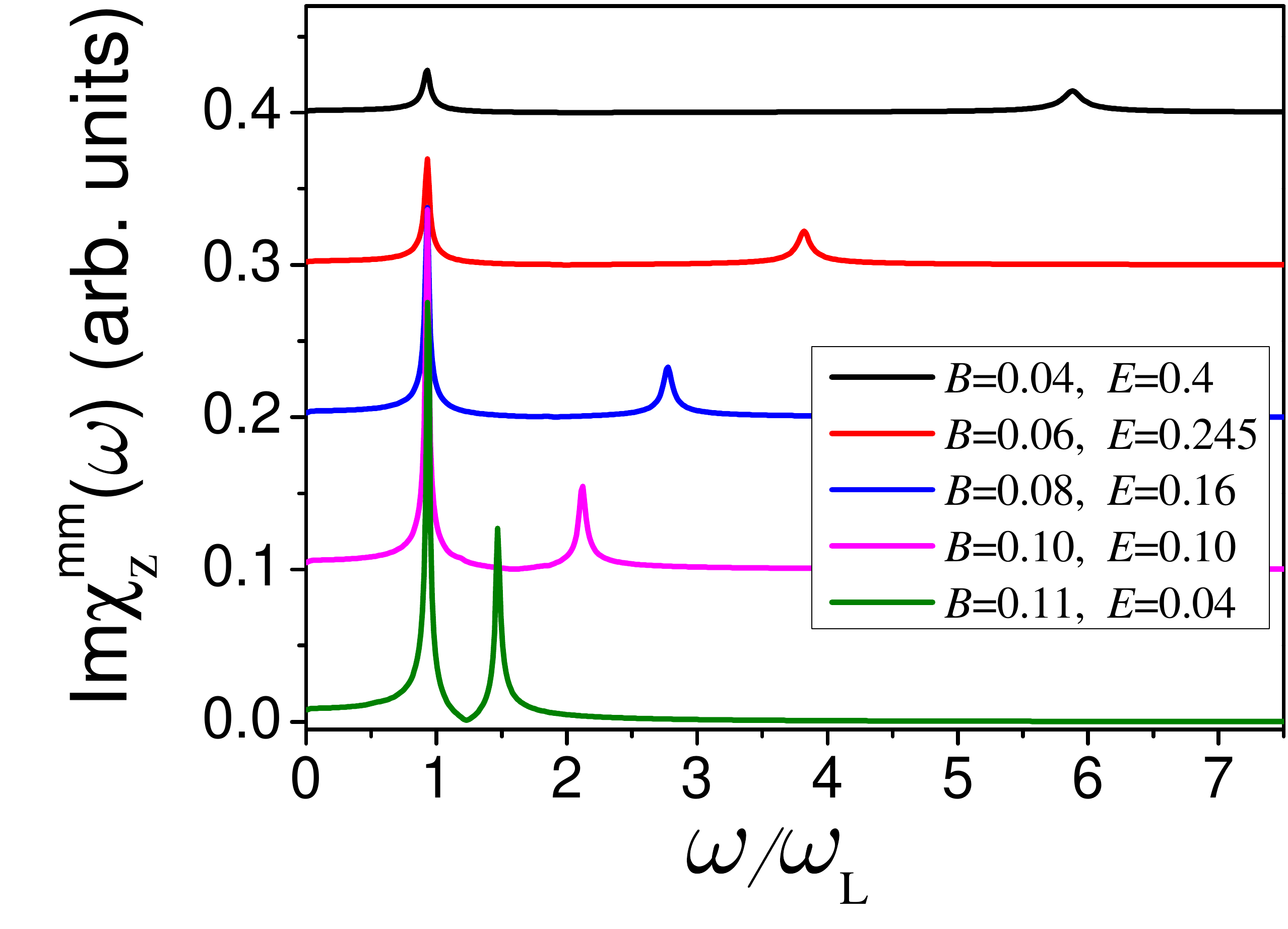}
\caption{\small
\textbf{The separation of peaks}.
The imaginary parts of the dynamical magnetic susceptibility
$\mathrm{Im}\protect\chi _{z}^{\mathrm{mm}}(\protect\omega )$ for various $B$ and $E$.
The electric field is tuned up such that the left peak (as a benchmark peak) is adjusted to
the same position for different magnetic field.
The curve is plotted with the applied electric field $E$ along an inclination angle $\protect\theta =%
\protect\pi /4$, and the frequency $\protect\omega $ is measured in unit of
$\protect\omega^{}_\mathrm{L}=\protect g\gamma _{e} B/\hbar$ with $B=0.1$. }
\label{fig:more}
\end{figure*}

Now we are in the position to perceive an entirely new strategy
for migratory birds' navigation.
We have witnessed the distinct feature of resonance-fluorescence spectrum
that is characterized by two parameters whose magnitudes rely on both
applied magnetic and electric fields.
If these two applied fields are the earth's magnetic field (geomagnetic field) and an internal bio-electric field
of a bird, then the aforementioned two parameters $\alpha$ and $\beta$
are related to the location and the self-effort of the bird.
As long as the frequency $\nu$ of the intrinsic driving field is assumed to
match the Zeeman energy of the geomagnetic field $B(x_\mathrm{b})$ at the breeding ground or birthland
(i.e., location $x_\mathrm{b}$),
the sensory spectrum of the bird at birthland possesses three peaks
satisfying double-resonance condition:
$\beta=\alpha=\alpha_0$. In Fig.~\ref{fig:strategy},
we adopted $\alpha^{}_0/\hbar=10.4564\protect\,\mathrm{rad/\upmu s}$
or equivalently $\alpha_0/h=1.6642\protect\,\mathrm{MHz}$ for $B(x_\mathrm{b})=58.0\mathrm{\upmu T}$,
a value of the geomagnetic field near $52^\circ$ latitude.

We know the value of the electric field
between two oppositely charged parallel surfaces
can be tuned simply by changing the surface-charge density
that depends on the curvature of the surfaces.
It is reasonable to suppose that the bird can change it's internal electric field (Fig.~\ref{fig:strategy}D).
At any place away from the bird's birthland,
the migratory bird is assumed to tune the parameter $\alpha$ of our model to match the single-resonance condition:
$\alpha=\alpha_0$ and $\alpha\neq\beta$,
with the help of its internal electric field
although the geomagnetic field at the location $x\neq x_\mathrm{b}$ does not match the resonance condition.
The single-resonant spectrum tuned up by the bird possesses five peaks (Fig.~\ref{fig:strategy}B)
and the splitting side-peak contains information of the location,
i.e.,
the more separation between the two side-peaks appears,
the farther away from its birthland it means (Fig.~\ref{fig:strategy}C and table S1).
Here the left one is benchmark peak and the spacing between the separate peaks can reflect the distance between its current location and destination.
Therefore  the bird can choose a right direction to move so that the two separate-peaks (Fig.~3C) are getting closer and closer.
As the frequency  of the intrinsic driving field is assumed to match
the Zeeman energy of the geomagnetic field at the birthland,
the sensory spectrum is the one under double-resonance condition once the bird arrives the birthland.

\subsection*{Disorientation caused by oscillating field}

\begin{figure*}[t]
\centering
\includegraphics[width=0.66\textwidth=0.66]{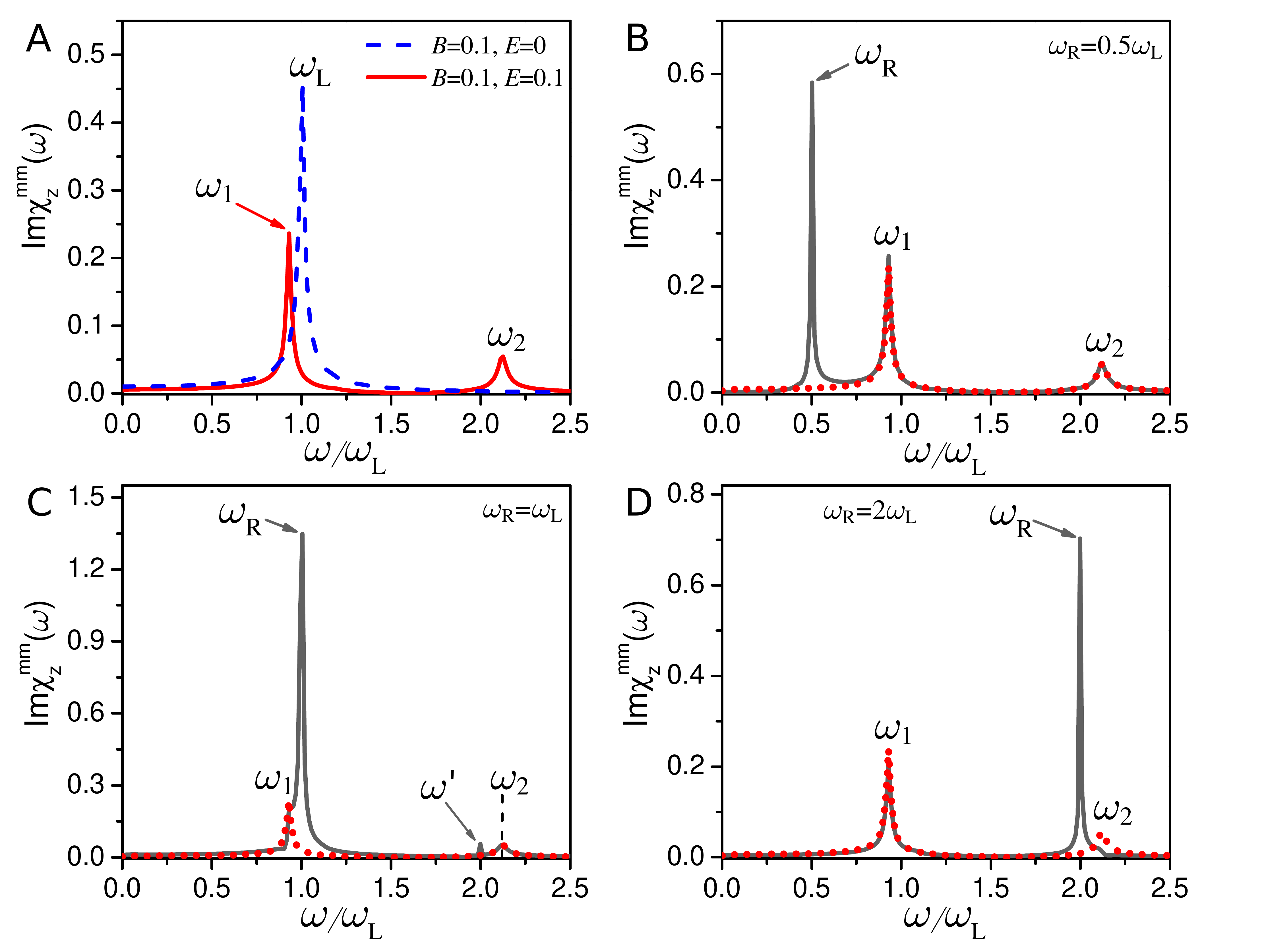}
\caption{\label{fig:disori}\small\sf
\textbf{Effects of added oscillating fields}.
(\textbf{A})
The single peak (blue dash-line) of Larmor frequency $\omega^{}_\mathrm{L}=g\gamma^{}_\mathrm{e}B/\hbar$
and the double-peaks (red line) $\omega_1$ and $\omega_2$ for navigation
are solved via simulation by letting $E=0$ and $E\neq0$ respectively.
Here $\gamma^{}_\mathrm{e}$ is the gyromagnetic ratio of the electron,
$g$ the g-factor of ferric,
and $B$ the geomagnetic field at the experiment site.
(\textbf{B}) to (\textbf{D})
The effects caused by the oscillating magnetic field of various radio frequencies
$\omega^{}_\mathrm{R}$, i.e.,
(B) $\omega^{}_\mathrm{R}=\omega^{}_\mathrm{L}/2$;
(C) $\omega^{}_\mathrm{R}=\omega^{}_\mathrm{L}$;
(D) $\omega^{}_\mathrm{R}=2\,\omega^{}_\mathrm{L}$.
Those black curves are the spectra disrupted by an oscillating magnetic field,
the red dot-lines (plotted for comparison) are the pure spectra without disrupt.
The navigation is based on measuring the difference between $\omega_1$ and $\omega_2$
after adjusting the benchmark peak $\omega_1$ to a standard position.}
\end{figure*}

In section~\ref{sec:LLG}, the separate peaks used in the navigation can also be calculated theoretically as the imaginary part of the dynamical magnetic susceptibility,
$\mathrm{Im}\,\chi^\mathrm{mm}_z$,
through a numerical simulation of LLG equation
by introducing a time-localized, uniform $\delta$-function pulse of magnetic field $B(t)=\delta(t)B$.
In terms of this numerical simulation, the effects caused by an oscillating magnetic field
can be studied conveniently.
Considering the experimental situations ever made by Ritz et al.~\cite{Ritz2009} several years ago,
we simulate the dynamics of the cluster in the presence of an additional oscillating magnetic field applied along the $y$-axis:
\begin{align*}
\boldsymbol{B}^{\prime }(t)=\allowbreak B_{1} \sin (\omega _{\mathrm{R}}t) \hat{y}.
\end{align*}
We then monitor time profiles of net spin $\Delta S_{z}(t)$ after applying a $\delta $-function pulse of magnetic field and calculate their Fourier transforms.
In Fig.~\ref{fig:disori}A, we show the calculated dynamical magnetic susceptibility
$\mathrm{Im}\chi _{z}^{\mathrm{mm}}(\omega )$
with (red line) and without (blue dash-line) electric field.

For those situations that had ever been observed
in the experiment~\cite{Ritz2009} on the disorientation caused by oscillating magnetic fields
of radio frequency $\omega^{}_\mathrm{R}$,
our simulation results are plotted in Figs.\,\ref{fig:disori}, B to D.
In our proposal of navigation strategy,
a key point is to adjust benchmark-peak $\omega_1$ to the right position
and then judge the separation of $\omega_2$ from it.
However,
the adjustable benchmark peak $\omega_1$ is disrupted in the case $\omega^{}_\mathrm{R}=\omega^{}_\mathrm{L}$ (Fig.~\ref{fig:disori}C)
while remains in the other two cases
$\omega^{}_\mathrm{R}=0.5\,\omega^{}_\mathrm{L}$
and
$\omega^{}_\mathrm{R}=2\,\omega^{}_\mathrm{L}$
(Figs.\,\ref{fig:disori}, B and D).
Additionally, a tiny peak $\omega'$ near $\omega_2$ appears
in that case (Fig.\,\ref{fig:disori}C), which also bewilders the bird's judgement for navigation.
Therefore, the bird's orientation ability is more significantly disrupted
when the radio frequency matches the Larmor frequency $\omega^{}_\mathrm{L}$, i.e., $\omega^{}_\mathrm{R}\simeq\omega^{}_\mathrm{L}$.
In that experiment~\cite{Ritz2009}, the test birds' orientation ability were significantly disrupted either by an oscillating magnetic field of frequency $1.315$ MHz
when geomagnetic field $B=46\, \upmu\mathrm{T}$, or by that of frequency $2.63$ MHz when $B=92\,\upmu\mathrm{T}$.
The corresponding Larmor frequency of the ferric sulfide cluster of our model
reads~\cite{SM} (see Eq.~S57),
$$\nu^{}_\mathrm{\small L}\;(\mathrm{in~MHz}) = 0.028693 B\;(\mathrm{in}~ \upmu\mathrm{T}).$$
It is evaluated to be $1.3199$ MHz when $B=46\, \upmu\mathrm{T}$
and to be $2.6398$ MHz when $B=92\,\upmu\mathrm{T}$,
clearly, they are much closer to the experimental data than the referenced~\cite{Ritz2009} Larmor frequencies of electrons: $1.288$ MHz and $2.576$ MHz.
Thus the result of the experimental observation~\cite{Ritz2009}
can be explained by our theory more precisely.

\section{Conclusion and outlook}

In conclusion, we have discovered theoretically a much more objective mechanism
that utilizes the separate peaks of resonance-fluorescence spectrum for navigation,
of which a distinctive forte is that the benchmark peak is naturally related to the geomagnetic field on the place of the migrating goal.
The separation of the side peaks contains the information of the distance between
the current location and the destination place of migratory bird,
and the change of the separation can guide the bird to choose the right direction to move.
Most importantly, the driven frequency in the model is related to the birthland,
the double-resonance peak appears without internal tuning electric field when the bird  arrives the destination.
Mouritsen~\cite{Mouritsen2018} addressed, recently,
twenty important mechanistic questions related to long-distance animal navigation that are expected to be solved over the next twenty years.
Clearly, several of them have been inspired by our present article.
The new sight we unveiled  is expected to arose more cues for readers to solve the other issues among those twenty mechanistic questions.
The numerical simulation via Landau-Lifshitz-Gilbert equation
enables us to
explain a previously observed disorientation effects caused by oscillating magnetic field
more clearly.
Thus the present work not only present an entirely new candidate mechanism for migratory navigation, but also motivates people to make
possible manmade practical devices or robots.
Actually, our theoretical approach opens up a new avenue of electrically tunable magnetic sensors.

\newpage

\newpage

\section*{Acknowledgments}

This work is supported by National Key R \& D Program of China, Grant No. 2017YFA0304304, and NSFC, Grant No. 11935012.

\newpage

\newpage

\newpage

\newpage

\newpage

\vspace{88mm}
\newpage

\end{document}